# Near-Field Optical Effect of a Core-Shell Nanostructure In Proximity to a Flat Surface

Wenping Cui[1], Mingda Li[2*] , Zuyang Dai[3], Qingping Meng[4] and Yimei Zhu[4]
[1]*Department of Physics, University of Bonn, 53113 Bonn, Germany*
[2]*Department of Nuclear Science and Engineering, Massachusetts Institute of Technology, Cambridge, MA 02139, USA*
[3]*Department of Physics and State Key Laboratory of Low-Dimension Quantum Physics, Tsinghua University, Beijing, 100084, China*
[4] *Department of Condensed Matter Physics, Brookhaven National Lab, Upton, NY11973, USA*

***Abstract.*** We provide an analytical solution to study the near field optical effect of a core-shell nanostructure in proximity to a flat surface, within quasi-static approximation. The distribution of electrostatic potential and field enhancement in this complex geometry are obtained by solving a set of linear equations. This analytical result can be applied to a wide range of systems associated with near field optics and surface plasmon polaritons. As an illustration of the power of this technique, we study the field attenuation effect of oxidized shell in a silver tip in near field scanning microscope. The thickness of oxidized layer can be monitored by measuring the light intensity. In addition, we propose a novel method to detect local temperature with spatial resolution down to nm scale, based on a Ag-Au core-shell structure.

[*]Corresponding author. E-mail: mingda@mit.edu

## 1. Introduction

The synthesis and studies of optical properties of composite core-shell nanostructures have attracted much attention [1-6]. The coupling between incident light and localized plasmon excitation generated on small particles leads to an enormous enhancement of local electric field [7, 8], which has been widely applied to surface enhanced Raman spectroscopies [9, 10] and Near-field Scanning Optical Microscopies (NSOM) [11, 12]. The scheme of the core-shell structure further allows a tunable optical response of the composite system, by adding additional degree of freedom of the shell composition and thickness. Xu *et al* [6] reported the synthesis of a $Fe_3O_4$-Au/Ag structure with tunable plasmonic resonant frequency, while Renteria-Tapia *et al* [13] measured the absorption spectra of Ag-$Ag_2O$ nanostructures grown on silica thin films.

Despite the blooming studies on synthesis and optical measurement of composite core-shell nanostructures, the theoretical description of such core-shell composite is still limited within the frame of Mie theory [5, 11], which is ideal to study light scattering from a composites core-shell nanostructure with spherical symmetry, but fails to describe the near-field optical properties when the composite sphere is above a flat surface, with no spherical symmetry.

A sphere in proximity to a surface has wide applications. In tip-enhanced Raman spectroscopy [10, 14] as well as NSOM [12], the tip-end can be well modeled by a sphere with finite curvature [12], thus can be well described within such geometry. Avarind and Metiu [15] provided an analytical solution of a single sphere close to a flat surface within quasi-static approximation, without considering the core-shell composite structure. To the best of our knowledge, the analytical solution of core-shell structure near a surface has not yet been reported.

In this study, using the trick of center-shifted sphere and bi-spherical coordinate [16], we are able to obtain an analytical solution for a core-shell structure in close proximity of a flat surface, within quasi-static approximation. This approximation is valid when the size of the core-shell is much smaller than the wavelength of incident light, thus can well describe the near-field behavior of core-shell nanostructure. Based on the solution, one can easily obtain the near-field optical properties, such as spatial distribution of potential and the local field enhancement, between a core-shell nanostructure and a nearby flat surface, by simply solving a set of linear equations. Since s-wave incident beam is proven to have much lower enhancing capability than p-wave incidence [15], here we consider the situation of monochromatic p-wave incidence only.

In the following we demonstrate the capability of our method in several material systems suitable for such a core-shell configuration. We found that in a common near-field device consisting of a tip, the increase of oxidation level of the silver tip will inevitably decrease the field enhancement, even if the oxidation layers (treated as shell, compared with the end of the tip as the core) have same thickness. On the other hand, since dielectric function changes as temperature varies, the resonant frequency at which the enhancement is peaked shifts as a function of local temperature. Therefore, the tip may be used as a component of thermometer to measure the local temperature of the flat surface, with a spatial resolution comparable to the size of the tip end.

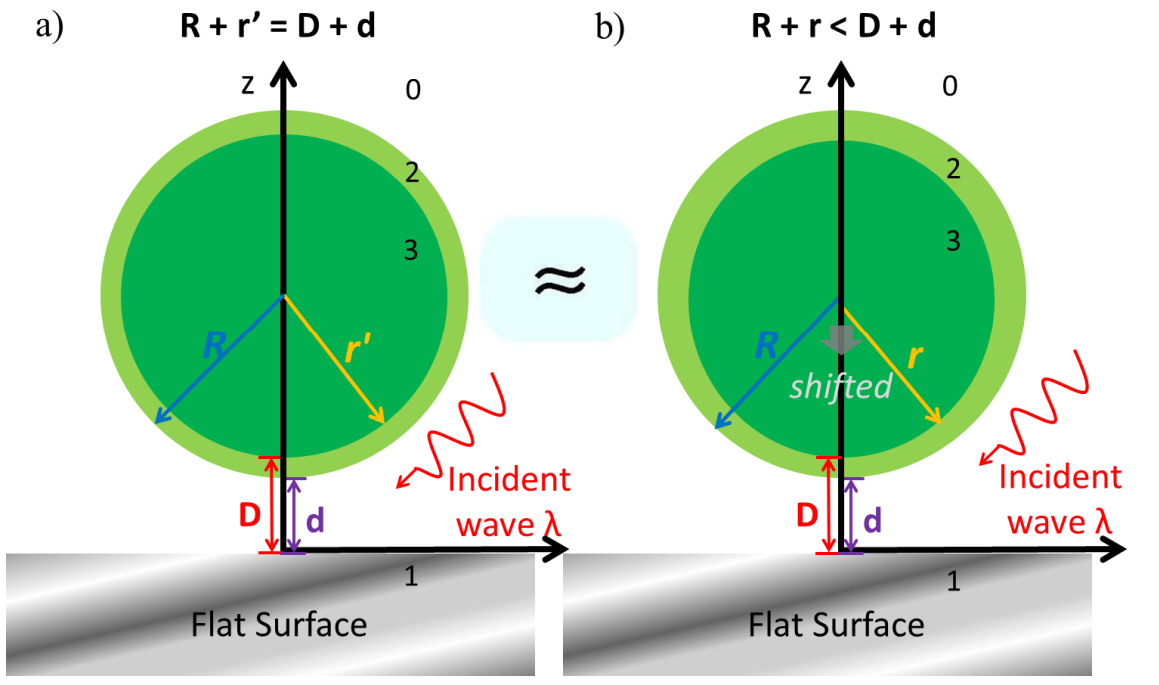

Fig. 1 The concentric core-shell structure(a) and non-concentric core-shell structure(b) in proximity to a flat surface. The latter has very similar near field optical properties at the gap between the sphere and the flat surface. When the outer shell and inner core shares the same foci point, the geometry in (b) becomes analytically solvable in Bi-spherical coordinate.

## 2. The Methodology for Calculating the Electric Field

### *2.1. The Geometrical Configuration*

The geometrical configuration for a film-shell-core structure is shown in Fig. 1(a). The outer-shell radius, inner-core radius, distance from shell-bottom to surface, distance from core-bottom to surface and incidence wavelength, are denoted as $R$, $r'$, $d$, $D$ and $\lambda$, respectively. We assume that $d+2R \ll \lambda$ valid, so that the retardation effect can be neglected and electric fields can be obtained by solving Laplace's equation $\nabla^2\Phi = 0$. The half-space in which the sphere is imbedded has a frequency independent dielectric constant of $\varepsilon_0$, which is the permeability of the media into which the system is immersed. For simplicity, we keep $\varepsilon_0$=1, treating the solvent as vacuum.

However, the Laplace's equation describing the configuration in Fig. 1(a) cannot be solved analytically. In order to grasp the essential feature and to simultaneously avoid the large-scale simulation, at least in the non-retardation region, the center of the inner core is slightly shifted with shrunk radius, keeping *D*, *d* and *R* as invariants. The equation for outer shell and inner core in Cartesian coordinate can be written as

$$x^2 + y^2 + \left(z - a\coth m_{1,2}\right)^2 = \frac{a^2}{\sinh^2 m_{1,2}} \qquad (1)$$

where $a \equiv \sqrt{(R+d)^2 - R^2}$ .

By defining $m_1 \equiv \operatorname{arccosh}\left(1 + d/R\right)$ for outer shell and $\sinh m_2 \equiv R\sinh\left(\operatorname{arccosh}\left(1+d/R\right)\right)/r$ for inner core, both spheres can then be simply denoted as ($\mu_{1,2}$, 0, 0) in bi-spherical coordinate, where the flat surface can be written as $\mu$=0. The geometrical configuration after shifting the boundary is illustrated in Fig. 1(b).

From this definition, it can be easily shown that $a\coth m_1 = R + d$ is valid, which means the center of the outer shell does not move at all. As to inner core, $D = a\coth m_2 - r$ and $r = a/\sinh m_2$ are valid. Since $m_2 > m_1$ and *coth* is a monotonic decreasing function, we obtain $D + r < d + R$ , that the center of the non-concentric inner core is lowered.

### 2.2. General Forms of Electric Potentials

With an incidence of *p*-polarization light, the electric potential for incidence wave can be written as

$$\Phi^i = -\mathbf{E}^i \cdot \mathbf{r} = -E_x^i x - E_z^i z \qquad (2)$$

where $E_x^i = -E_0\cos q_0$ , $E_y^i = 0$ and $E_z^i = E_0\sin q_0$ , $q_0$ is the incidence angle with respect to *x*-axis, $E_0$ is the magnitude of the incident electric field.

After the reflection by the flat surface, the potential for the reflected wave is written as

$$\Phi^r = -\mathbf{E}^r \cdot \mathbf{r} = -E_x^r x - E_z^r z \qquad (3)$$

where $E_x^r = E_0 r_p \cos q_0$ , $E_y^r = 0$ and $E_z^r = E_0 r_p \sin q_0$ , and $r_p \equiv \dfrac{e_1/e_0\cos q_0 - a}{e_1/e_0\cos q_0 + a}$ is the *p*-wave reflection coefficient with $a \equiv \sqrt{e_1/e_0 - \sin^2 q_0}$ .

Similarly, for transmitted wave into flat surface (region 1), the potential is written as

$$\Phi^t = -\mathbf{E}^t \cdot \mathbf{r} = -E_x^t x - E_z^t z \qquad (4)$$

where $E_x^t = -(1 - r_P)E_0\cos q_0$ , $E_y^t = 0$ , and $E_z^t = \dfrac{e_0}{e_1}E_0\left(r_P + 1\right)\sin q_0$ .

#### 2.2.1. Electric Potential in Region 0: inert medium

In a bi-spherical coordinate, inert medium region 0 satisfies $0 < m < m_1$ . The total potential in this region can be written as

$$V_0 = \Phi^i + \Phi^r + V_0^s \qquad (5)$$

where $V_0^s \equiv V_0^s(m,h,j)$ is the generic form of scattering potential induced by the core-shell structure:

$$V_0^s(m,h,j) = F\sum_{n\ge|m|}^{\infty}\sum_{m=-\infty}^{+\infty}\left(A_n^m e^{(n+1/2)m} + B_n^m e^{-(n+1/2)m}\right)Y_n^m(\cos h, j) \qquad (6)$$

where $F \equiv \sqrt{\cosh m - \cos h}$ , $A_n^m$ and $B_n^m$ are expansion coefficients, $Y_n^m$ are spherical harmonics defined in Appendix A.

#### 2.2.2. Electric Potential in Region 1: Flat Surface

The surface *z*=0 in Cartesian coordinate separates the space as $m > 0$ upper space and $m < 0$ lower space, which is denoted as region 1. The total potential in this region can be written as

$$V_1 = \Phi^t + V_1^s \qquad (7)$$

where $\Phi^t$ is the electric potential of transmitted wave when the core-shell structure is absent, $V_1^s$ is the potential induced by the core-shell structure.

$$V_1^s(m,h,j) = F\sum_{n\ge|m|}^{\infty}\sum_{m=-\infty}^{+\infty} D_n^m e^{+(n+1/2)m} Y_n^m(\cos h, j) \qquad (8)$$

In this region there is no expansion coefficients linked with $e^{-(n+1/2)m}$ to prevent potential blow-up since $m < 0$ .

#### 2.2.3. Electric Potential in Region 2: Outer Shell

Region 2 satisfies $m_1 < m < m_2$ , with potential

$$V_2(m,h,j) = F\sum_{n\geq|m|}^{\infty}\sum_{m=-\infty}^{+\infty}\left(G_n^m e^{(n+1/2)m} + H_n^m e^{-(n+1/2)m}\right)Y_n^m(\cosh,j) \quad (9)$$

*2.2.4. Potential in Region 3: Inner Shell*

Region 3 satisfies $m > m_2$ with potential

$$V_3(m,h,j) = F\sum_{n\geq|m|}^{\infty}\sum_{m=-\infty}^{+\infty} C_n^m e^{-(n+1/2)m} Y_n^m(\cosh,j) \quad (10)$$

In this region there is no expansion coefficients linked with $e^{+(n+1/2)m}$ to prevent potential blow-up approaching to the focal point where $m \to +\infty$ .

## *2.3. Calculation of Electric Field*

The potentials $V_0 - V_3$ can be calculated by requiring the boundary conditions of Maxwell's equations:

$$V_1(m=0^-,h,j) = V_0(m=0^+,h,j) \quad (11a)$$

$$V_0(m=m_1,h,j) = V_2(m=m_1,h,j) \quad (11b)$$

$$V_2(m=m_2,h,j) = V_3(m=m_2,h,j) \quad (11c)$$

$$e_0\frac{\partial V_0(m=0,h,j)}{\partial m} = e_1(w)\frac{\partial V_1(m=0,h,j)}{\partial m} \quad (12a)$$

$$e_0\frac{\partial V_0(m=m_1,h,j)}{\partial m} = e_2(w)\frac{\partial V_2(m=m_1,h,j)}{\partial m} \quad (12b)$$

$$e_2(w)\frac{\partial V_2(m=m_2,h,j)}{\partial m} = e_3(w)\frac{\partial V_3(m=m_3,h,j)}{\partial m} \quad (12c)$$

In order to match the boundary conditions, potentials $\Phi^i$ , $\Phi^r$ and $\Phi^t$ have to be rewritten in a bi-spherical coordinate. This can be achieved by expansion [16]:

$$x = -(\cosh m - \cosh)^{1/2}\sqrt{8}a\cos j\sum_{n=1}^{\infty}P_n^1(\cosh)e^{-(n+1/2)m} \quad (13a)$$

$$z = (\cosh m - \cosh)^{1/2}\sqrt{2}a\sum_{n=0}^{\infty}(2n+1)P_n^0(\cosh)e^{-(n+1/2)m} \quad (13b)$$

Using this method, the expansion coefficients at different regions in eqs (6-10) are coupled through a system of linear equations, which are the main result of this study and shown in Appendix B due to its lengthy coefficients. In numerical calculation of the coefficients, terms with $|m|\leq 1$ survive in the linear equations, with n is cut off at $n$=20 to ensure charge neutrality and convergence.

Based on the expansion coefficients, the electric field can be further calculated numerically from the potential gradient in Cartesian coordinate. The transformation from Cartesian coordinate to bi-spherical coordinate is shown in Appendix A eq. A1.

## 3. Results and Discussions

In principle, this method can be applied to any core-shell system in proximity to a flat surface, to understand the near-field enhancement effect, as long as the dielectric function of each region is known. Fig. 2 shows a general pattern of spatial distribution of total electric potential $V_0$ and the intensity enhancement is defined as $I/I_{inc} \equiv |E|^2 / |E_0|^2$, where a silver core is encrusted by a silver oxide shell on a silver surface. The incident angle is set to be 45° and the total electric Field $E$ is obtained numerically from total potential in Fig.2 (a). This configuration can describe the field attenuation yield from the oxidation layer in a silver-tip based plasmonic device.

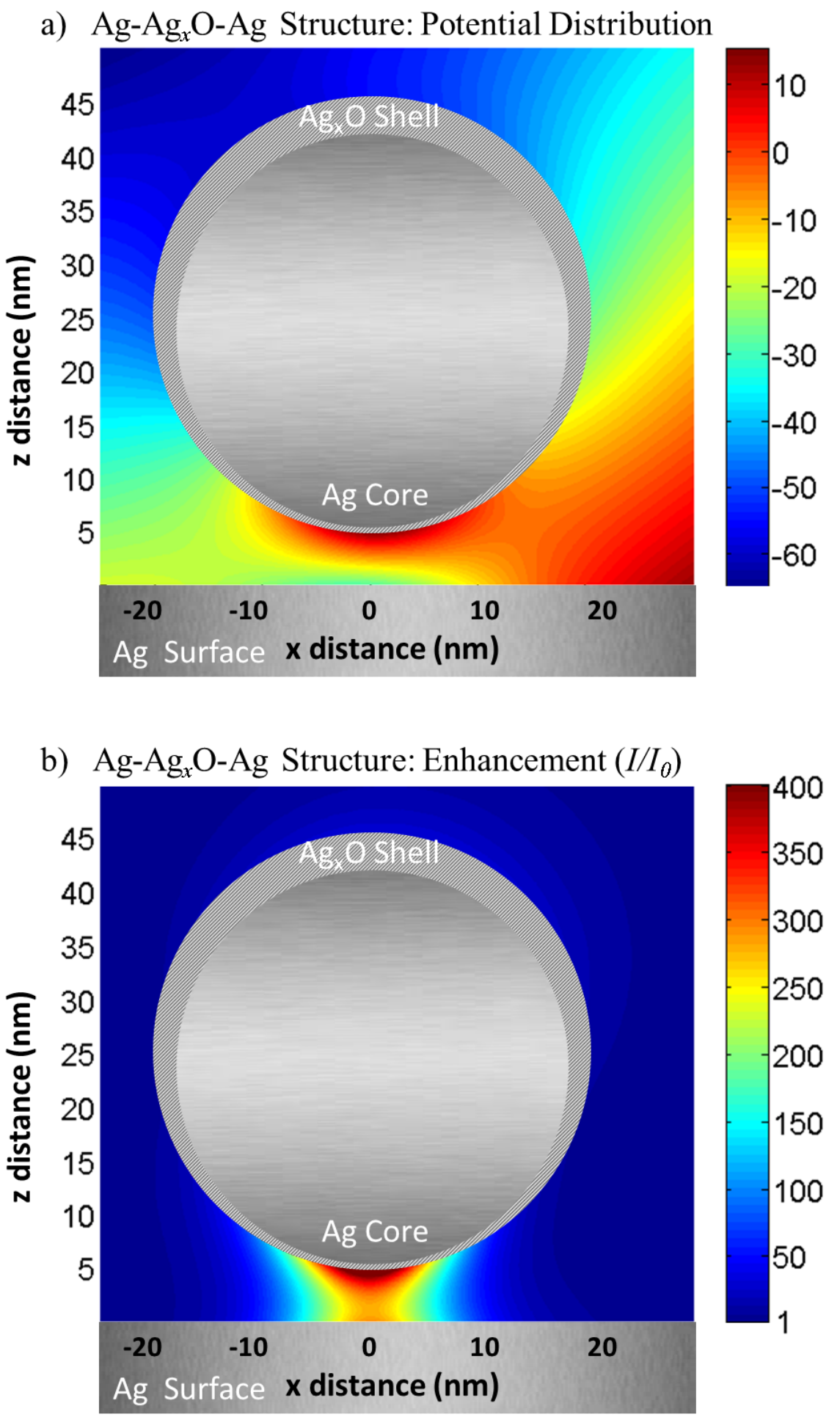

Fig. 2 Spatial distribution of electric potential (a) and local intensity enhancement (b) of Ag core-Ag$_x$O shell-Ag surface composite structure. In (a), the magnitude of incidence electric field $E_0$ is set to be unity. The frequency here is 3.2eV, with oxygen flux 3sccm (See in Sec 3.1), shell radius $R$=20nm, core radius $r$=18nm and the distance from shell bottom to surface $d$=5nm. The region of maximum intensity enhancement is localized near the sphere in the gap between the shell and the film.

### *3.1. Near Field Attenuation caused by Oxidation Layer*

In order to study the near field effect of shell oxidation, the frequency dependent dielectric function of silver oxide is substituted into eqs. (B3-B8) as an input parameter. We adopt the dielectric functions of silver oxides reported in [17], where the dielectric functions of $Ag_xO$ films were studied by ellipsometry at different levels of oxygen flux. The flat surface is chosen as gold [18]. We found that even oxide shells are assumed to have the same thickness, the local intensity attenuation at various oxygen flux behaves differently, which is illustrated in Fig. 3.

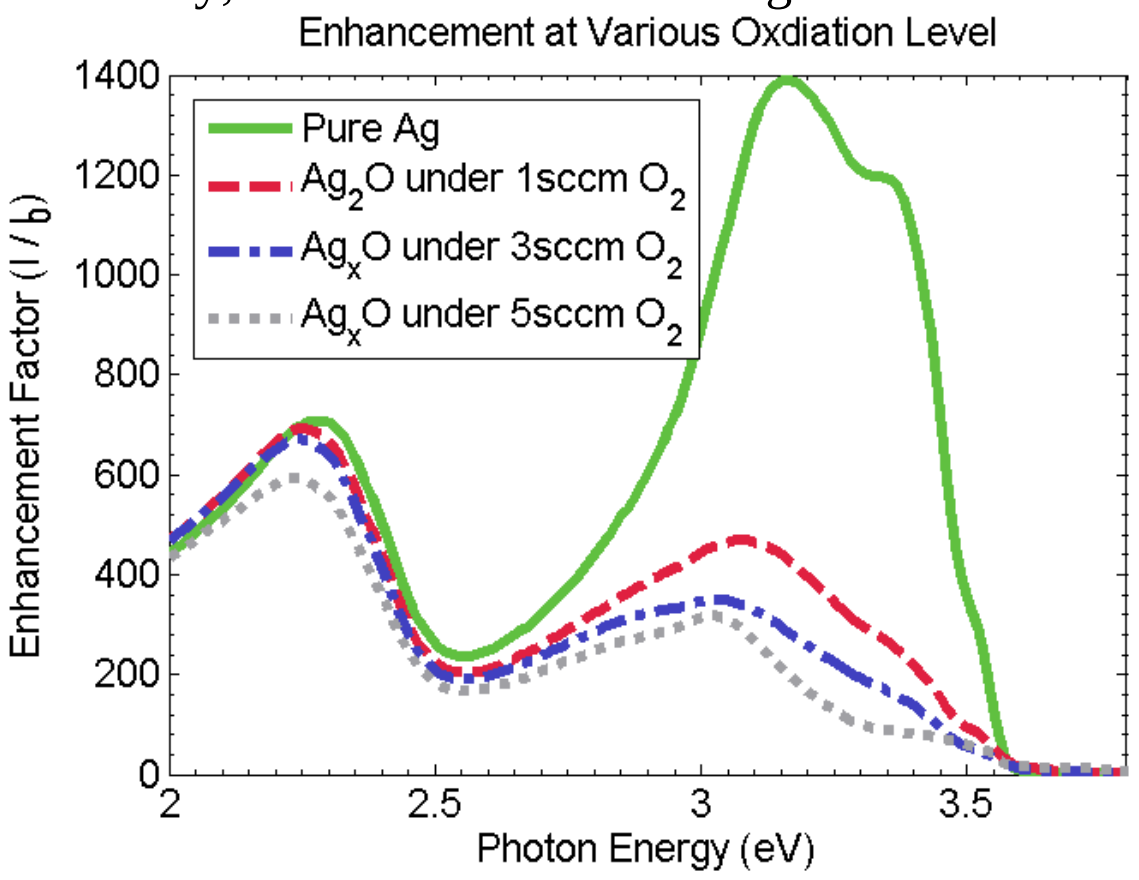

Fig. 3 Maximum enhancement factor as a function of photon energy, at various oxygen flow for oxidation $Ag_xO$ layer. Different curves share the same thickness of $Ag_xO$ layer, but with distinct dielectric functions due to different oxygen flow during the oxidation process. The geometry here is $D$=20nm, $r$=16nm and $d$=2nm. The peak at 3.1eV and 2.2 eV correspond to the resonance peak of Ag and Au, respectively.

The green solid-line denotes near field spectrum of pure silver tip without any oxidation. At 1sccm oxygen flow (red dash line), the oxidation layer dramatically reduces the peak intensity. When the oxygen flow increases, the peak intensity monotonically decreases, but stabilizes at 3 and 5 sccm $O_2$ level (blue dot-dash curve and grey dotted curve). This is quite reasonable since it correctly simulates the behavior of degradation caused by oxidation as well as oxidation saturation, at the beginning of the reaction.

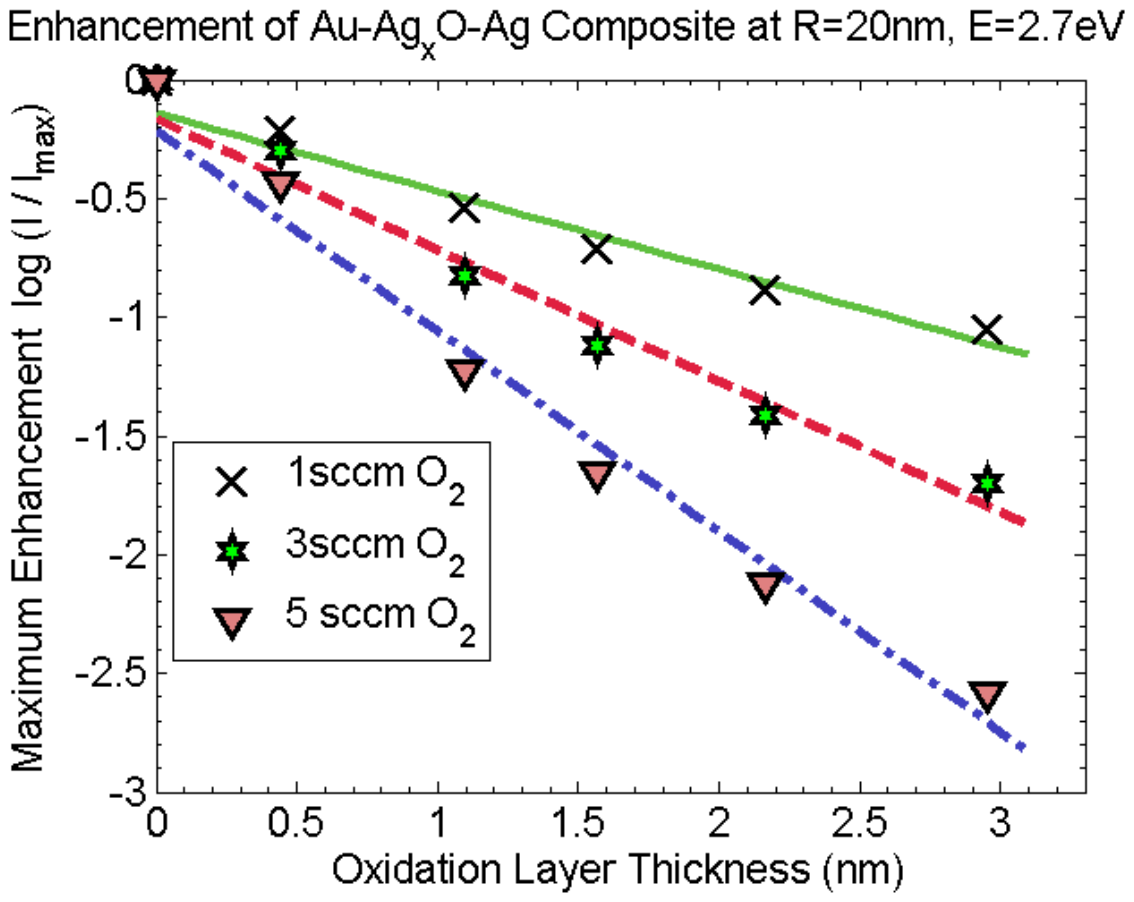

Fig. 4. The logarithm of normalized intensity as a function of $Ag_xO$ layer thickness, at various oxygen flux levels. The geometries at thickness $D$-$d$ of 0.43nm are the same as Fig.3, while other points are obtained by keeping $R$ and $d$ constants with varying $D$-$d$. The coefficient of determination $R^2$ for linear fitting are 0.963, 0.969 and 0.976, respectively. The linear relationship between normalized intensity and shell thickness sets up a simple way to determine the thickness of oxidized shell, at least at the beginning of the oxidation reaction.

When the oxidation layer starts growing, the intensity drops accordingly, as shown in Fig.4. The logarithm of intensity normalized with respect to the pure-Ag intensity changes linearly with layer thickness. This can be understood from the optical tunneling effect of localized plasmon [19, 20]. The enhancement ~3eV is caused by the coupling between localized plasmon at the Ag-core and incident light. While with the existence of thin oxidation layer, which performs as a tunneling barrier, the optical transmission coefficient is reduced exponentially. This further reduces the coupling strength and results on the excited intensity. Therefore, by measuring the relative intensity during an oxidation process, the oxidation layer growth can be monitored in-situ, at least at the initiative stage of the reaction, where the intensity is still strong enough to be measured. Moreover, providing that the temporal information of oxidation time is given, when oxygen flux and layer thickness are known, the atomic oxygen reaction rate with silver surface can be further obtained [21]. A similar situation with silver flat surface is illustrated in Appendix Fig. C1.

### *3.2. Local temperature Sensor with ultra-high spatial resolution based on Ag-Au Core-Shell Composite*

Dielectric function is generally a function of temperature [22-27], hence a change of the near field enhancement, such as resonant frequency shift or intensity variation, may be utilized as an indicator of temperature in a highly localized region. Fig.5 provides an example of such frequency shift of Ag core-Au shell- GaAs surface structure. The temperature dependence of GaAs film is obtained from [27].

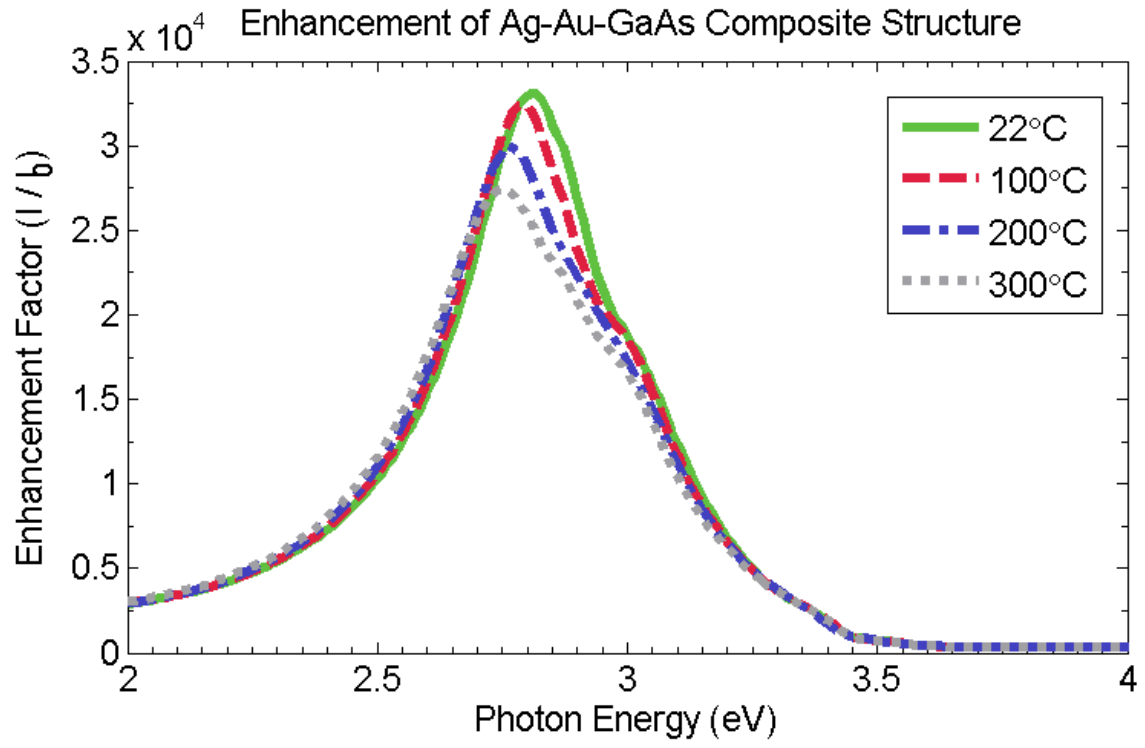

Fig. 5 Intensity enhancement of Ag-Au core-shell structure above a GaAs flat surface. The geometrical configuration is $R$=20nm, $d$=1nm, $D$=1.5nm. The photo energy with maximum enhancement factor shifts as a function of temperature.

This result might not be surprising from first glimpse since the bandgap GaAs has a temperature dependent dielectric functions (Fig. C2). However, this is only achieved with the existence of a thin shell. The comparison of resonant energy between pure Ag structure and Ag-Au core shell structure is illustrated in Fig. 6. The former case can be calculated by setting both core and shell with same dielectric function of Ag. Without the Au shell, resonance of the pure Ag structure does not vary much (blue-dashed line) due to a fixed plasmon resonance of silver. However, with carefully engineered Au-shell thickness, the linear energy shift can be realized. When the shell thickness continues to grow, the Au-shell contribution dominates the near-field behavior and the resonant frequency becomes fixed again, See in Appendix Fig. C3. This can be understood as a difference of resonance energy between silver and gold.

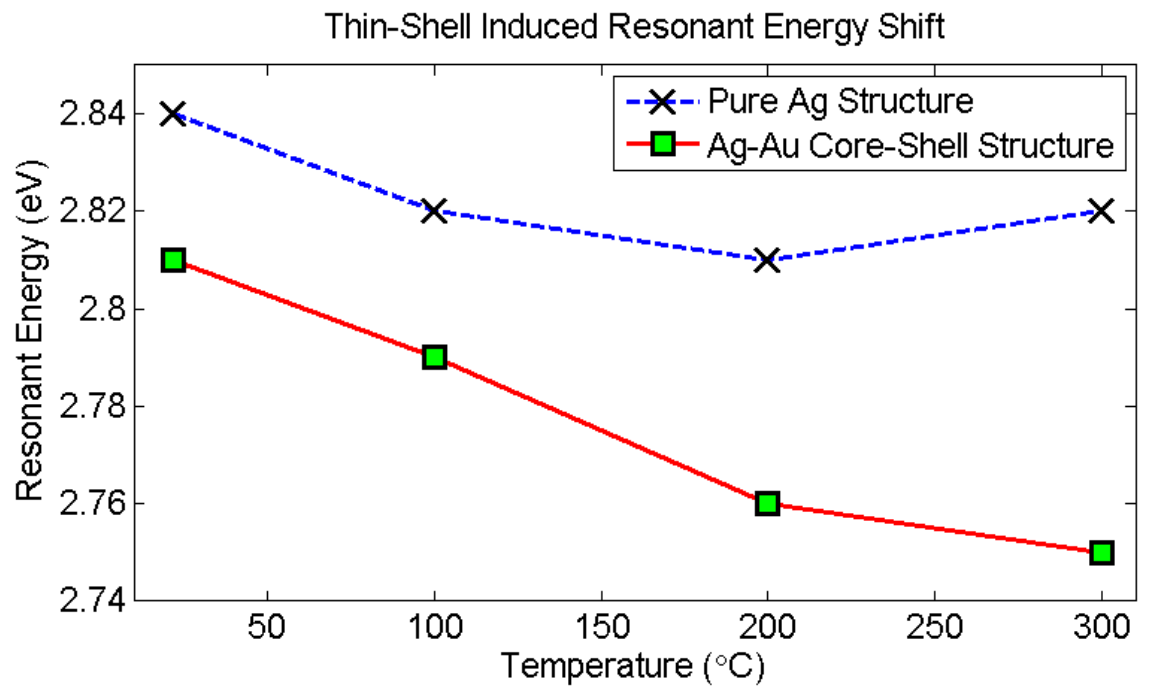

Fig. 6 The resonance frequency as a function of temperature, for pure Ag structure and Ag-Au core shell structure. Both Situations share the same geometry as described in Fig.4. The linear relation ($R^2$=0.96) between resonant frequency and temperature laid the foundation for the application local temperature measurement.

When a core-shell structure is assembled, the degree of freedom of shell thickness allows more flexibility to fine tune the trend of resonance frequency as the function of surface temperature. The almost linearity of resonant frequency as temperature in the case of Ag-Au core-shell structure can then be utilized as an indicator of local temperature. Since the light intensity is enhanced in a very localized region, this configuration might have application as thermometer with around 10nm spatial resolution, if the based temperature of certain material surface can be calibrated.

## 4. Conclusions

In this work, we compute the light enhancement effect of a spherical core-shell structure on a flat surface under the incidence of p-polarization light. This configuration can be used to model the near-field enhancement of a tip with realistic oxidation conditions, when it is placed in proximity to a surface, and extract the thickness of the oxidized layer of the tip directly. We also propose the use of a tip as a thermal sensor to measure nanoscale temperature of the underlying material. With this approach, the complexity of near field behavior of core-shell nanocomposite is reduced to solve a set of linear equations with a given geometry and dielectric functions, leading to achieve better near-field optical performance of nanodevices by engineering the core-shell structure.

## Acknowledges

Author Mingda Li would like to thank Prof. Ju Li for his generous support and helpful discussions. Mingda Li also would like to thank Prof. Yonggang Li, Prof. Jianming Jia and Prof. Hongyi Li for their kind suggestions.

**Appendix A: Useful Formula in Deriving the Main Results**

1. Transform from Cartesian to Bi-spherical Coordinate:

$$(x,y,z)=\left(\frac{a\sinh h\cos j}{\cosh m-\cos h},\frac{a\sinh h\sin j}{\cosh m-\cos h},\frac{a\sinh m}{\cosh m-\cos h}\right), \tag{A1}$$

where $a=\sqrt{(d+R)^2-R^2}$, $m\in(-\infty,+\infty)$, $h\in[0,p]$, $j\in[0,2p)$.

2. Inverse transform:

$$(m,h,j)=\left(\sinh^{-1}\left(\frac{2az}{Q}\right),\cos^{-1}\left(\frac{R^2-a^2}{Q}\right),\tan^{-1}\left(\frac{y}{x}\right)\right), \tag{A2}$$

where $a=\sqrt{(d+R)^2-R^2}$, $Q\equiv\sqrt{(R^2+a^2)^2-(2az)^2}$, $R\equiv\sqrt{x^2+y^2+z^2}$

3. Properties of Associated Legendre Polynomials and Spherical Harmonics:

$$Y_n^m(x,j)=-\sqrt{\frac{2n+1}{4p}}\sqrt{\frac{(n-m)!}{(n+m)!}}P_n^m(x)e^{imj} \tag{A3}$$

$$P_n^{-1}(x)=(-1)^1\frac{(n-1)!}{(n+1)!}P_n^1(x) \tag{A4}$$

$$Y_n^{-1}(x,j)=-\sqrt{\frac{2n+1}{4p}}\sqrt{\frac{(n-1)!}{(n+1)!}}P_n^1(x)e^{-ij} \tag{A5}$$

$$Y_n^{+1}(x,j)=\sqrt{\frac{2n+1}{4p}}\sqrt{\frac{(n-1)!}{(n+1)!}}P_n^{+1}(x)e^{+ij} \tag{A6}$$

4. Equations to ensure charge neutrality and criteria for convergence:

$$\sum_{n=0}^{\infty}(2n+1)P_n^0(\cos h)=0 \tag{A7}$$

$$\sum_{n=1}^{\infty}(n+1/2)P_n^1(\cos h)=0 \tag{A8}$$

5. Recursion formula:

$$\int_{-1}^{+1}P_m^0(x)P_n^0(x)dx=\frac{2}{2n+1}d_n^m \tag{A9}$$

$$\int_{-1}^{+1}P_m^1(x)P_n^1(x)dx=\frac{2n(n+1)}{2n+1}d_n^m \tag{A10}$$

$$\int_{-1}^{+1}xP_m^0P_n^0dx=\frac{1}{2n+1}\int_{-1}^{+1}\left[nP_{n-1}^0+(n+1)P_{n+1}^0\right]P_m^0dx=\frac{2n}{4n^2-1}d_m^{n-1}+\frac{2(n+1)}{(2n+1)(2n+3)}d_m^{n+1} \tag{A11}$$

$$\int_{-1}^{+1}xP_m^1P_n^1dx=\frac{1}{2n+1}\int_{-1}^{+1}\left[(n+1)P_{n-1}^1+nP_{n+1}^1\right]P_m^1dx=2\frac{n(n+1)(n-1)}{(2n+1)(2n-1)}d_m^{n-1}+\frac{2n(n+1)(n+2)}{(2n+1)(2n+3)}d_m^{n+1} \tag{A12}$$

**Appendix B: Main Results (Linear equations to Solving expansion Coefficients)**

Define $F_0 \equiv \sqrt{2}E_0 a$ , $F \equiv F(m) = \sqrt{\cosh m - \cosh h}$ , $F_0^+ = F_0 \sin q_0 (1 + r_P)$ , $F_0^- = F_0 \cos q_0 (1 - r_P)$ , $X_1 = \dfrac{e_0 - e_1(w)}{e_0 + e_1(w)}$, $X_2 = \dfrac{e_2(w) - e_3(w)}{e_2(w) + e_3(w)}$. The relations for the expansion coefficients can be written as:

$$Z_n^{-1} = -Z_n^1, \quad Z = A, B, C, D, G, H \tag{B1}$$

$$Z_n^m = 0, \quad Z = A, B, C, D, G, H \quad m \neq 0, \pm 1 \tag{B2}$$

$$D_n^m = A_n^m + B_n^m, \quad B_n^m = X_1 A_n^m, \quad C_n^m = G_n^m e^{(2n+1)m_2} + H_n^m, \quad m = 0, \pm 1 \tag{B3}$$

For m=0 components

$$-F_0^+ \sqrt{4p\,(2n+1)} + A_n^0 e^{(2n+1)m_1} + X_1 A_n^0 = G_n^0 e^{(2n+1)m_1} + H_n^0 \tag{B4}$$

$$\begin{aligned}
&\left[(2n+1)\cosh m_2 + X_2 \sinh m_2\right] G_n^0 - \sqrt{\frac{2n+3}{2n+1}}(n+1)e^{m_2} G_{n+1}^0 + X_2 \sqrt{\frac{2n+3}{2n+1}}(n+1)e^{-(2n+2)m_2} H_{n+1}^0 \\
&+ X_2\left[\sinh m_2 - (2n+1)\cosh m_2\right] e^{-(2n+1)m_2} H_n^0 - \sqrt{\frac{2n-1}{2n+1}} n e^{-m_2} G_{n-1}^0 + X_2 \sqrt{\frac{2n-1}{2n+1}} n e^{-2nm_2} H_{n-1}^0 = 0
\end{aligned} \tag{B5}$$

$$\begin{aligned}
&2F_0^+ e^{-(2n+1)m_1}\left[(n+1)e^{-m_1} - n e^{+m_1}\right] = \left[\sinh m_1\left(1 + X_1 e^{-(2n+1)m_1}\right) + (2n+1)\left(1 - X_1 e^{-(2n+1)m_1}\right)\cosh m_1\right]\sqrt{\frac{2n+1}{4p}} A_n^0 \\
&- \left[e^{+m_1} - X_1 e^{-(2n+2)m_1}\right](n+1)\sqrt{\frac{2n+3}{4p}} A_{n+1}^0 - \left[e^{-m_1} - X_1 e^{-2nm_1}\right] n \sqrt{\frac{2n-1}{4p}} A_{n-1}^0 \\
&- \frac{e_2}{e_0}\left[(2n+1)\cosh m_1 + \sinh m_1\right]\sqrt{\frac{2n+1}{4p}} G_n^0 + \frac{e_2}{e_0}\left[(2n+1)\cosh m_1 - \sinh m_1\right]\sqrt{\frac{2n+1}{4p}} e^{-(2n+1)m_1} H_n^0 \\
&+ \frac{e_2}{e_0}(n+1)\sqrt{\frac{2n+3}{4p}}\left[e^{+m_1} G_{n+1}^0 - e^{-(2n+2)m_1} H_{n+1}^0\right] + \frac{e_2}{e_0} n \sqrt{\frac{2n-1}{4p}}\left[e^{-m_1} G_{n-1}^0 - e^{-2nm_1} H_{n-1}^0\right]
\end{aligned} \tag{B6}$$

For m=1 components

$$+A_n^1 e^{(2n+1)m_1} + X_1 A_n^1 - G_n^1 e^{(2n+1)m_1} - H_n^1 = F_0^- \sqrt{\frac{4p\,n(n+1)}{2n+1}} \tag{B7}$$

$$\begin{aligned}
&X_2\left[\sinh m_2 - (2n+1)\cosh m_2\right] e^{-(2n+1)m_2} H_n^1 + X_2 \sqrt{\frac{(2n+1)(n^2-1)}{2n-1}} e^{-2nm_2} H_{n-1}^1 + X_2 \sqrt{\frac{(2n+3)(n^2+2n)}{2n+1}} e^{-(2n+2)m_2} H_{n+1}^1 \\
&+ \left[(2n+1)\cosh m_2 + X_2 \sinh m_2\right] G_n^1 - \sqrt{\frac{(2n+1)(n^2-1)}{2n-1}} e^{-m_2} G_{n-1}^1 - \sqrt{\frac{(2n+3)(n^2+2n)}{2n+1}} e^{+m_2} G_{n+1}^1 = 0
\end{aligned} \tag{B8}$$

$$
\begin{aligned}
-2e^{-(2n+1)m_1}F_0^- \sinh m_1 &= \left[\left(1+X_1 e^{-(2n+1)m_1}\right)\sinh m_1 + (2n+1)\left(1-X_1 e^{-(2n+1)m_1}\right)\cosh m_1\right]\sqrt{\frac{2n+1}{4p\,(n^2+n)}}A_n^1 \\
&-\left[e^{+m_1} - X_1 e^{-(2n+2)m_1}\right]\sqrt{\frac{(2n+3)(n+2)}{4p\,(n+1)}}A_{n+1}^1 - \left[e^{-m_1} - X_1 e^{-2nm_1}\right]\sqrt{\frac{(2n-1)(n-1)}{4p\,n}}A_{n-1}^1 \\
&-\frac{e_2}{e_0}\sqrt{\frac{2n+1}{4p\,n(n+1)}}\left[(2n+1)\cosh m_1 + \sinh m_1\right]G_n^1 + \frac{e_2}{e_0}\sqrt{\frac{(2n+3)(n+2)}{4p\,(n+1)}}\left[e^{+m_1}G_{n+1}^1 - e^{-(2n+2)m_1}H_{n+1}^1\right] \\
&+\frac{e_2}{e_0}\sqrt{\frac{(2n-1)(n-1)}{4p\,n}}\left[e^{-m_1}G_{n-1}^1 - e^{-2nm_1}H_{n-1}^1\right] + \frac{e_2}{e_0}\sqrt{\frac{2n+1}{4p\,n(n+1)}}\left[(2n+1)\cosh m_1 - \sinh m_1\right]e^{-(2n+1)m_1}H_n^1
\end{aligned}
\tag{B9}
$$

## Appendix C

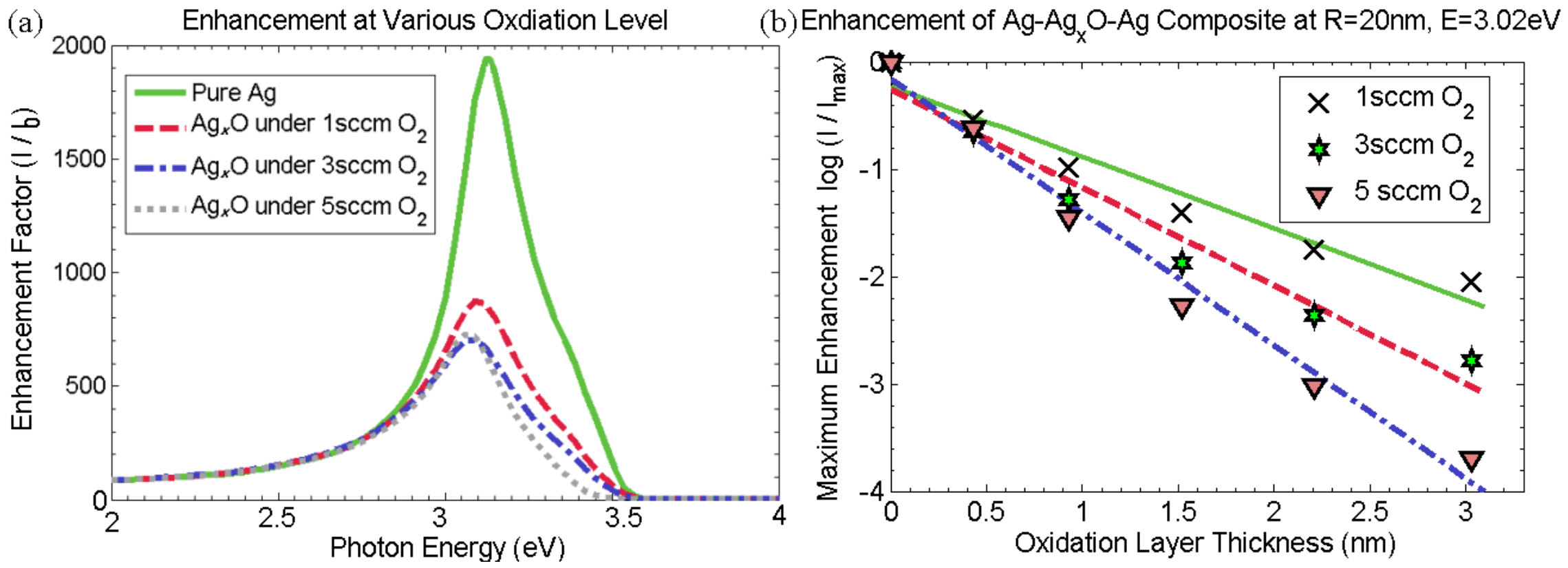

Fig. C1 Enhancement spectrum of Ag-Ag$_x$O-Ag composite structure (a) and normalized intensity at E=3.02eV (b), at various oxidation level. In (a), the geometry is the same as Fig.2, in (b), ***R*** and ***d*** are fixed with varying ***D-d***. Linear relationships of logarithm of intensity with oxidation layer thickness again appear, with coefficient of determination 0.953, 0.961 and 0.984, respectively.

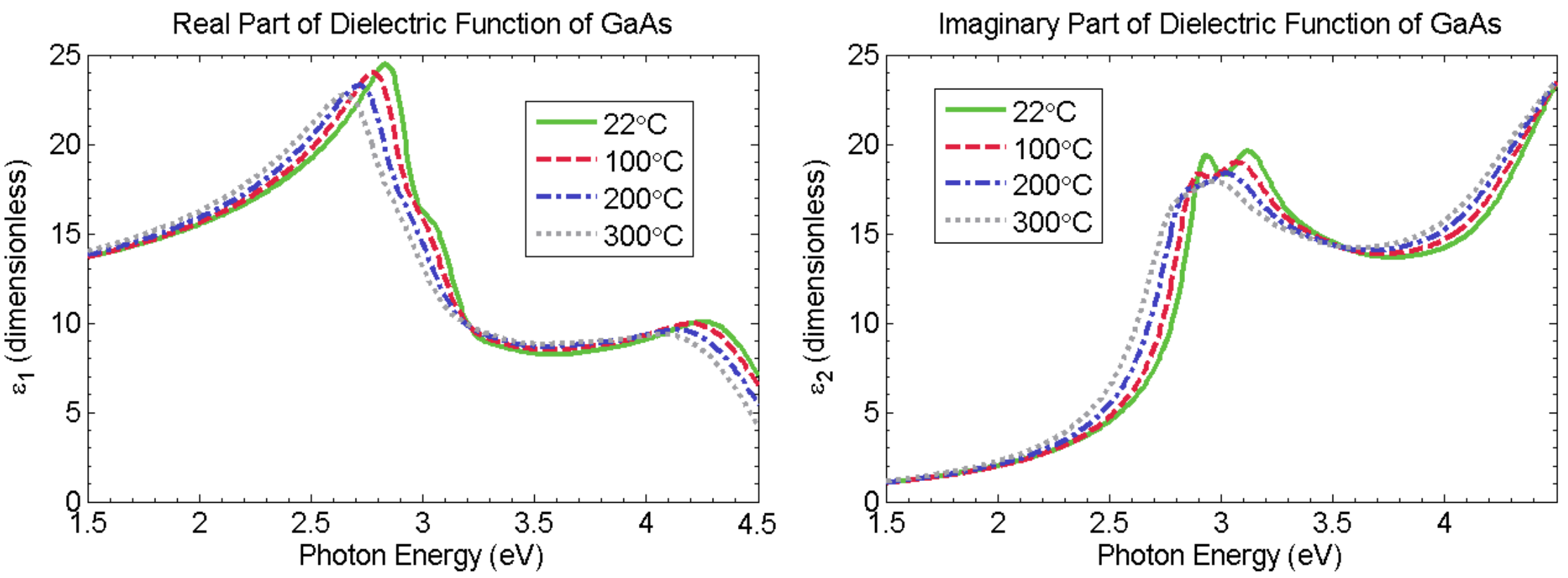

Fig. C2 Real and Imaginary part of Dielectric functions of GaAs as a function of temperature. Despite the small shift of dielectric functions, it is this shift combined with silver tip that determines the final resonance frequency.

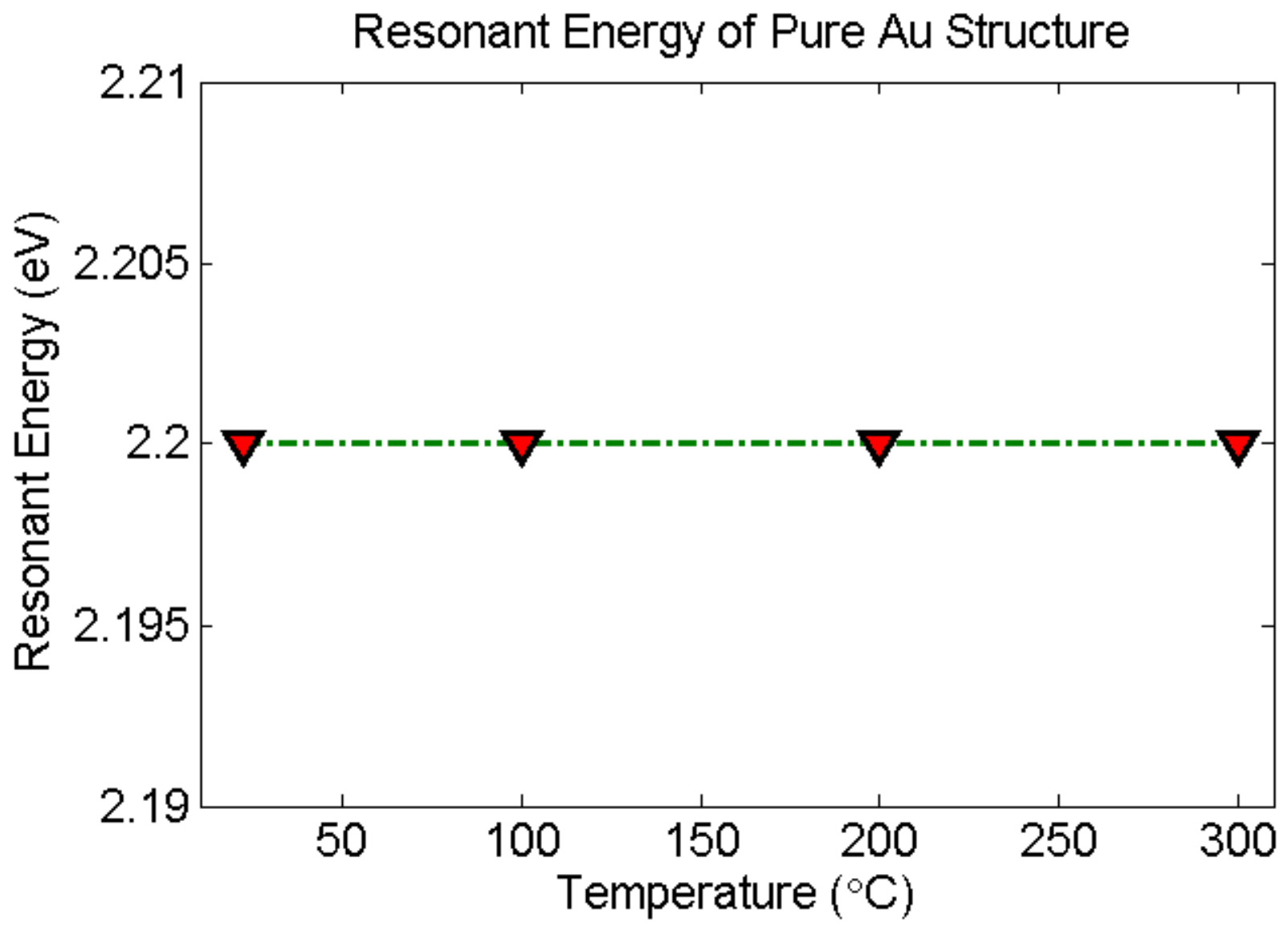

Fig. C3 The inner core and outer shell are both gold here, as opposed to Fig.5, In this case, the resonant function does not shift at all. This can be understood from Fig. A1, that at $E_{Au}$=2.2eV, the temperature variation of GaAs surface is small, compared with resonant frequency of silver $E_{Ag}$=2.82eV, where the difference of dielectric function of GaAs at different temperature is significant. Therefore micro-engineering of Ag-Au structure has the potential to fine tune the resonance energy to a band where the dielectric function of the underlying material has the most temperature sensitivity.